\documentclass[aps,prb,twocolumn,superscriptaddress,showkeys,showpacs]{revtex4-2}
\usepackage[T1]{fontenc}
\usepackage{amsmath}
\usepackage{graphicx}
\usepackage{tikz-cd}
\usepackage{eso-pic}

\begin{document}

\title{Direct observation of the electronic structure of even-layer puckered SnTe monolayer films on graphene}

\author{Satoru Ichinokura}
\email{ICHINOKURA.satoru@nims.go.jp}
\affiliation{Center for Basic Research on Materials, National Institute for Materials Science, Tsukuba, 305-0003, Japan}

\author{Shunsuke Tsuda}
\affiliation{Center for Basic Research on Materials, National Institute for Materials Science, Tsukuba, 305-0003, Japan}

\author{Kiyohisa Tanaka}
\affiliation{UVSOR Synchrotron Facility, Institute for Molecular Science, Okazaki, 444-8585, Japan}

\author{Koichiro Yaji}
\affiliation{Center for Basic Research on Materials, National Institute for Materials Science, Tsukuba, 305-0003, Japan}
\affiliation{Unprecedented-scale Data Analytics Center, Tohoku University, Sendai, 980-8578, Japan}
\affiliation{Photon Science Innovation Center (PhoSIC), Sendai, 980-8572, Japan}

\begin{abstract}

We report a direct observation of the electronic structure of monolayer SnTe on graphene by angle-resolved photoemission spectroscopy. By combining low-energy electron diffraction, momentum microscopy, core-level spectroscopy, and first-principles calculations, we show that the observed monolayer-limit electronic structure is consistent with a puckered even-layer biatomic-layer building block rather than a nonpuckered square-flat structure. The observed band dispersion and its polarization dependence are reproduced by a puckered structure with a substrate-renormalized buckling amplitude. We further show that graphene actively interacts with ultrathin SnTe by selecting its in-plane orientational texture, renormalizing the buckling amplitude, and modifying the local interfacial electrostatic environment in the monolayer limit. In comparison, multilayer SnTe exhibits a more ringlike momentum-space distribution, interlayer-split valence bands, and a stronger $p$-type character, revealing the onset of thickness evolution from an interface-stabilized puckered ultrathin limit toward a more bulklike state.

\end{abstract}

\maketitle

\AddToShipoutPictureFG*{%
  \AtPageLowerLeft{%
    \raisebox{3mm}[0pt][0pt]{%
      \makebox[\paperwidth][r]{%
        \parbox[b]{0.48\paperwidth}{%
          \raggedleft
          \fontsize{7.2}{8.6}\selectfont
          \color{black!45}
          \textcopyright\ 2026 American Physical Society. This is the accepted manuscript of:\\
          S. Ichinokura et al., Phys. Rev. B 114, 165408 (2026).\\
          The final published version is available at\\
          \texttt{https://doi.org/10.1103/p3hj-vdc2}.
        }%
        \hspace*{10mm}%
      }%
    }%
  }%
}

Group-IV monochalcogenides and related ultrathin chalcogenide films, represented by MX (M = Ge, Sn; X = S, Se, Te), have emerged as an important class of two-dimensional materials because reduced dimensionality stabilizes low-symmetry puckered structures and amplifies symmetry-breaking electronic responses. In this family, the puckered distortion is a key ingredient that drives in-plane ferroelectricity, orbital anisotropy, and a variety of symmetry-enabled functionalities, as demonstrated experimentally in monolayer SnS, SnSe, and ultrathin SnTe \cite{Higashitarumizu2020NatCommun_SnSInPlaneFE,Chang2020NanoLett_SnSeDomains,Chang2016Science_SnTeFerroelectricity,Chang2019AdvMater_EnhancedPolarizationSnTe}. These symmetry-broken lattices further provide a fertile setting for spin-orbit-coupled and nonlinear phenomena, including Rashba-type spin splitting, Berry-curvature dipoles, shift currents, and ultrafast polarization control \cite{Su2021MatDes_RashbaSnSSnSe,Kim2019NatCommun_BerryCurvatureSnTe,Lee2020APL_RashbaSnTeThinFilm,Slawinska2020_2DMater_SpintronicsSnTe,Jin2024npjComputMater_ShiftCurrentSnTe,Shin2020npjComputMater_NonlinearPhononicsSnTe}. Related chalcogenide systems such as GeTe have also highlighted how ferroelectricity can be directly coupled to spin texture control \cite{Krempasky2018PRX_OperandoSpinTextureGeTe}.

Bulk SnTe, by contrast, adopts a high-symmetry rocksalt structure and, owing to its small band gap, is prone to band inversion and the emergence of topological crystalline insulator (TCI) phases, as established on both single crystals and thin films \cite{Tanaka2012NatPhys_TCIinSnTe,Yan2014SurfSci_SnTeSi111,Guo2014APLMat_PbSnTeSTO001,Zhang2017JESRP_SnTe111}.
Indeed, if the lattice symmetry is retained down to the monolayer limit, SnTe(001) has been predicted to realize a two-dimensional TCI \cite{Liu2014NatMater_2DTCISnTe,Liu2015NanoLett_CrystalFieldTCI}. In practice, however, ultrathin SnTe does not generically remain in this idealized square-flat limit. Rather, as in other monochalcogenide systems, a puckered structure built from biatomic layers is energetically favored in the two-dimensional limit \cite{Chang2019AdvMater_EnhancedPolarizationSnTe,Chang2019APLMat_GrowthPhaseSnTeGraphene}. Such a puckered phase is more naturally realized as a larger-gap, symmetry-broken ferroelectric semiconductor and therefore lies outside the simple band-inverted TCI picture \cite{Fu2019PRB_SubstrateEffectsSnTe,Kim2019NatCommun_BerryCurvatureSnTe,Lee2020APL_RashbaSnTeThinFilm}.

Recent work has shown that this balance can be shifted by the substrate. In particular, trilayer-equivalent SnTe on NbSe$_2$ has been reported to enter a two-dimensional topological crystalline insulating regime under substrate-induced compressive strain \cite{Jing2026NatCommun_2DTCISnTe}. By contrast, graphene interacts much more weakly with SnTe and has been shown by scanning tunneling microscopy studies to support ultrathin SnTe in the ferroelectric limit with layered antipolar ordering and domain formation \cite{Chang2016Science_SnTeFerroelectricity,Chang2019APLMat_GrowthPhaseSnTeGraphene,Chang2019AdvMater_EnhancedPolarizationSnTe,Chang2019PRL_StandingWavesSnTeML}. However, direct determination by angle-resolved photoemission spectroscopy (ARPES) of whether the graphene-supported ultrathin limit realizes a puckered, symmetry-broken electronic structure rather than the idealized nonpuckered picture invoked in simple two-dimensional TCI scenarios has remained unavailable.

In this work, we directly determine the momentum-resolved electronic structure of monolayer-limit SnTe on graphene. By combining ARPES with low-energy electron diffraction (LEED), momentum microscopy, core-level spectroscopy, and first-principles calculations, we show that the graphene-supported ultrathin limit realizes a puckered even-layer biatomic-layer building block rather than the idealized nonpuckered square-flat structure invoked in simple two-dimensional TCI scenarios. We further show that graphene is not a passive support: it selects the in-plane orientational texture, modifies the lattice parameters relevant to the monolayer limit, and changes the local interfacial electrostatic environment. Finally, by comparing monolayer and multilayer films, we identify the onset of thickness evolution from an interface-stabilized puckered ultrathin limit toward a more bulklike regime.

Ultrathin SnTe films were grown in situ on epitaxial graphene formed on an n-type 4H-SiC(0001) substrate. The in-plane symmetry and reciprocal-space pattern were examined by LEED at an electron energy of 74 eV. Constant-energy maps were measured by momentum microscopy using a 10.9 eV vacuum-ultraviolet laser source \cite{Yaji2024eJSSNT_PhotoemissionMicroscopy,Yaji2024STAMMethods_iSPEM,Tsuda2024eJSSNT_ToFPEEM}. Valence-band ARPES and Te $4d$/Sn $4d$ core-level spectra were measured at BL5U of UVSOR. Electronic-structure calculations were performed within density-functional theory (DFT) using fully relativistic projector-augmented-wave pseudopotentials and the Perdew--Burke--Ernzerhof generalized gradient approximation. Further experimental and computational details are given in the Supplemental Material \cite{SupplementalMaterial}.



\begin{figure}[t]
\includegraphics[width=\columnwidth,trim=10 20 560 0,clip]{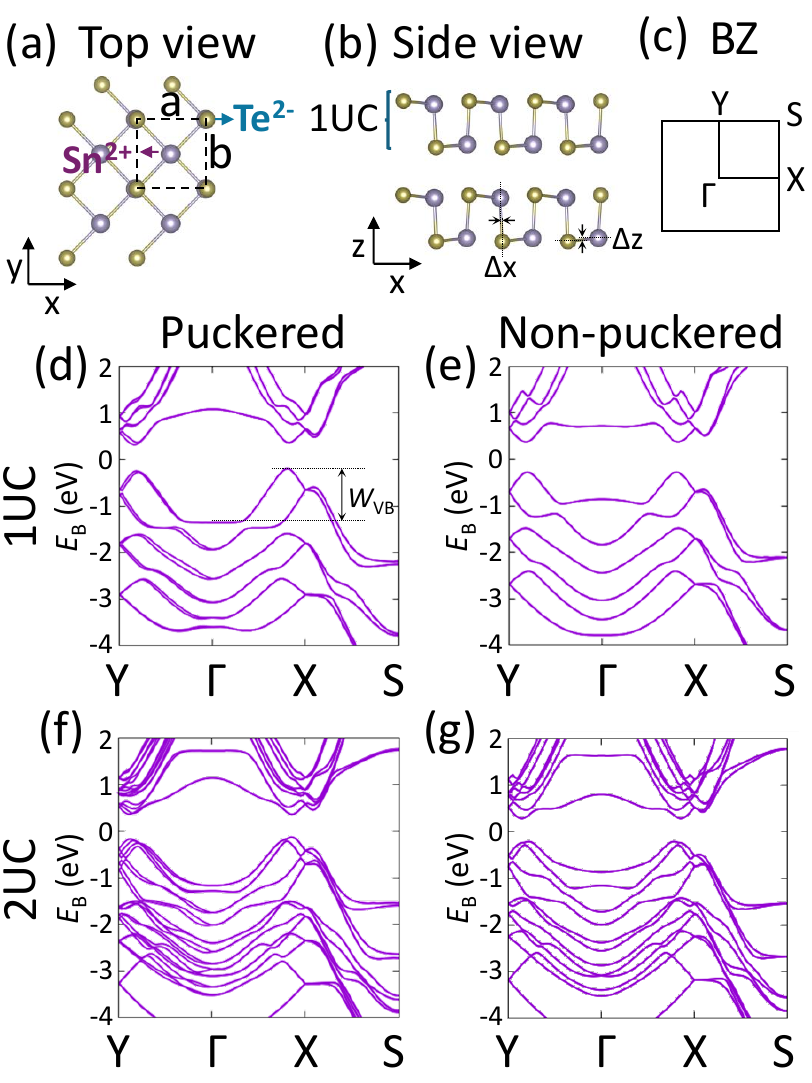}
\caption{
\label{fig:fig1}
Structural and electronic reference models for ultrathin SnTe.
(a),(b) Top and side views of the 2UC puckered structure.
(c) Corresponding pseudotetragonal Brillouin zone.
(d),(e) Calculated band structures of puckered and nonpuckered 1UC SnTe.
(f),(g) Calculated band structures of puckered and nonpuckered 2UC SnTe.
}
\end{figure}

Figure~1 introduces the structural and electronic reference models used in this work. Figures~1(a) and 1(b) show the top and side views of the puckered structure, and Fig.~1(c) shows the corresponding pseudotetragonal Brillouin zone. Here, 1UC and 2UC denote one- and two-unit-cell-thick SnTe slabs, respectively.

Using the relaxed 1UC puckered structure with $a=4.560$~\AA\ and $b=4.530$~\AA, we obtain finite in-plane displacement and buckling, indicating that the puckered structure is energetically favored.
For comparison, the nonpuckered reference calculations were performed with the averaged in-plane lattice constant, $a=b=4.545$~\AA.
The calculated band structures of the puckered and nonpuckered 1UC models are shown in Figs.~1(d) and 1(e), respectively. Compared with the puckered case, the nonpuckered square-flat limit loses the weakly anisotropic valence-band features.

The 2UC results are shown in Figs.~1(f) and 1(g). In the puckered 2UC model, interlayer coupling increases the number of valence bands and produces a ladderlike structure near $\Gamma$. The structural parameters and calculated quantities are summarized in Table~I. At both 1UC and 2UC thicknesses, the puckered structure is lower in energy than the nonpuckered one, and the valence-band width differs markedly between the two cases.

\begin{table}[t]
\caption{\label{tab:fig1models}
Calculated structural parameters and band characteristics for the 1UC and 2UC SnTe models. $\Delta x$: in-plane Sn--Te displacement; $\Delta z$: local buckling amplitude; $W_{\mathrm{VB}}$: valence-band width; $\Delta E$: relative energy from the nonpuckered model at the same thickness.}
\begin{ruledtabular}
\begin{tabular}{lccccc}
Layer & Structure & $\Delta x$ (\AA) & $\Delta z$ (\AA) & $\Delta E$ (meV/f.u.) & $W_{\mathrm{VB}}$ (eV) \\
\hline
1UC & Puckered  & 0.10 & 0.29 & -20.2 & 1.153 \\
1UC & Non-puck.     & 0    & 0    & 0     & 0.590 \\
2UC & Puckered  & 0.15 & 0.22 & -17.6 & 1.035 \\
2UC & Non-puck.     & 0    & 0    & 0     & 0.660 \\
\end{tabular}
\end{ruledtabular}
\end{table}


\begin{figure}[t]
\includegraphics[width=\columnwidth,trim=10 0 370 0,clip]{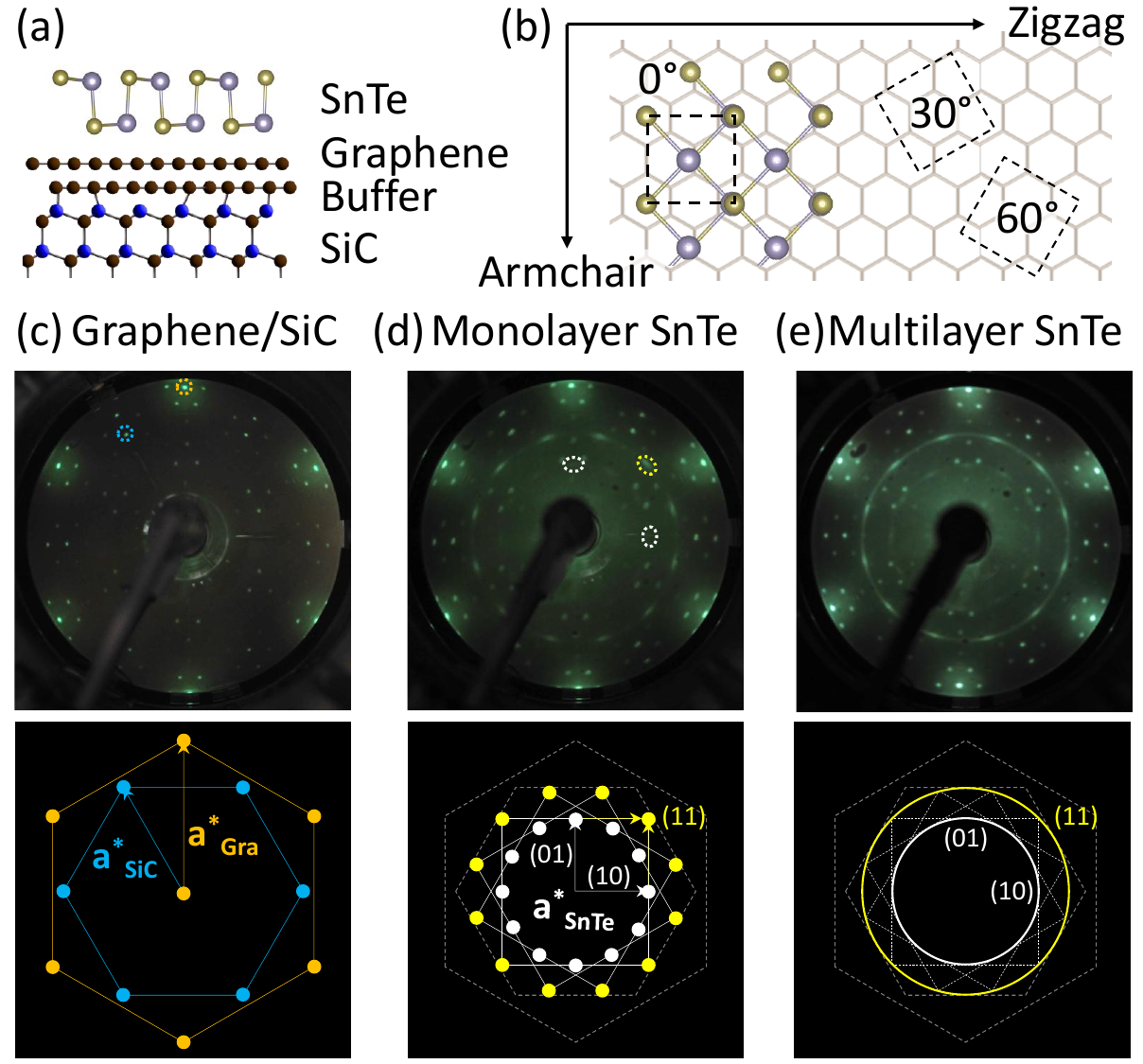}
\caption{
\label{fig:fig2}
LEED characterization of graphene-supported ultrathin SnTe.
(a) Schematic side view of the sample structure consisting of SnTe on graphene/buffer/SiC.
(b) In-plane orientation relationship between the pseudotetragonal SnTe lattice and the graphene lattice, illustrating domains rotated by 0$^\circ$, 30$^\circ$, and 60$^\circ$.
(c)--(e) LEED patterns of Graphene/SiC, monolayer SnTe, and multilayer SnTe, respectively, together with schematic reciprocal-space patterns shown below.
}
\end{figure}

Figure~2 summarizes the structural characterization of graphene-supported ultrathin SnTe by LEED. During molecular beam epitaxy (MBE) growth, SnTe was formed on graphene/buffer/SiC, as shown schematically in Fig.~2(a). The graphene/SiC substrate [Fig.~2(c)] exhibits the expected sixfold reciprocal-space features of graphene and SiC. After SnTe deposition, the monolayer film shows discrete SnTe-derived diffraction features, including the (01) and (11) spot arrays, that form an apparent 12-fold pattern [Fig.~2(d)]. This is naturally explained by a superposition of pseudotetragonal domains whose principal axes are locked to the graphene lattice with relative rotations of 0$^\circ$, 30$^\circ$, and 60$^\circ$ [Fig.~2(b)].

From the LEED patterns, the in-plane lattice constant of SnTe is estimated to be $4.49\pm0.02$~\AA\ for the monolayer film and $4.44\pm0.02$~\AA\ for the multilayer film. These values are consistent with previous reports for monolayer SnTe and with the Te--Te distance of bulk cubic SnTe \cite{Chang2016Science_SnTeFerroelectricity}, and suggest a thickness-dependent relaxation toward a more bulklike in-plane lattice scale. In the multilayer [Fig.~2(e)], the same basic diffraction geometry is preserved but the intensity becomes much more ringlike, indicating a broader orientational spread and/or increased mosaicity. Thus, ultrathin SnTe on graphene is not a single-domain crystal but a graphene-constrained multidomain pseudotetragonal film, with the orientational locking strongest in the monolayer limit.

\begin{figure}[t]
\includegraphics[width=\columnwidth,trim=18 40 438 20,clip]{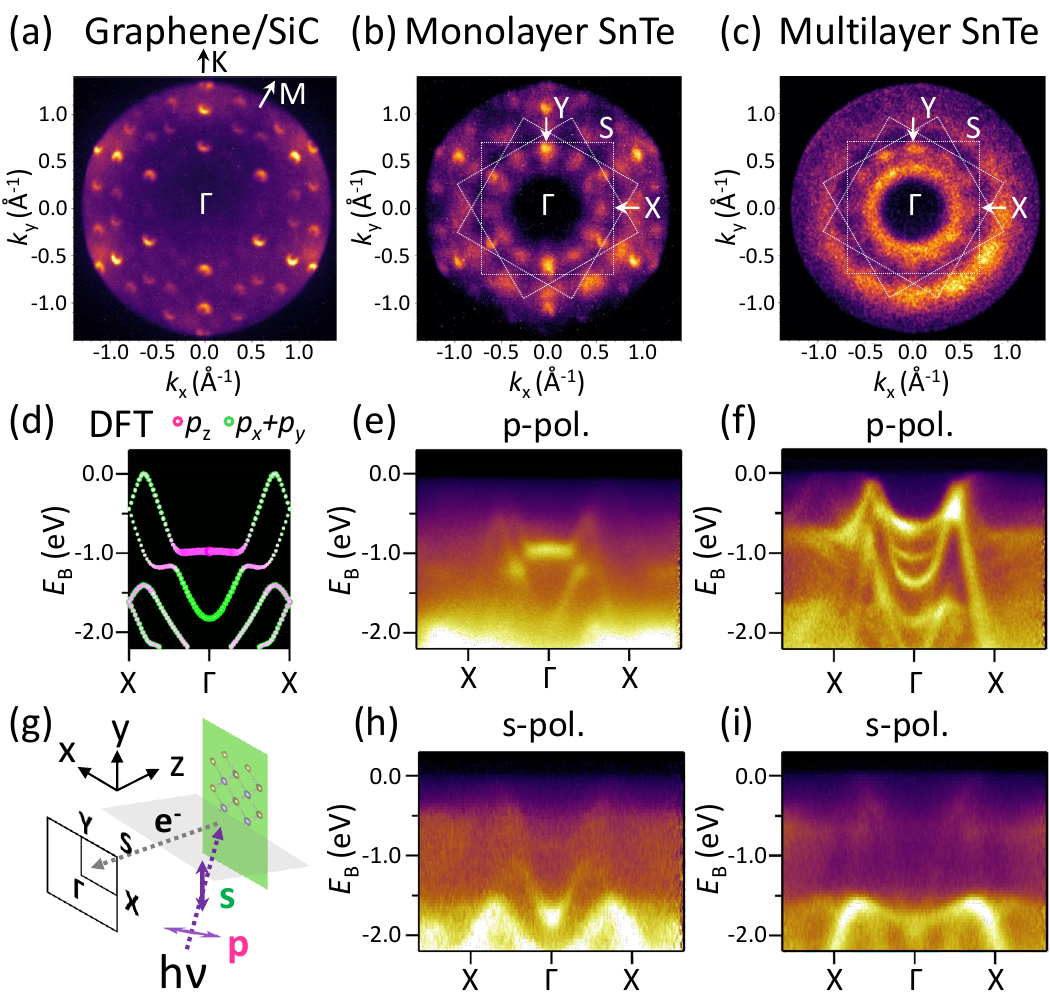}
\caption{
\label{fig:fig3}
Momentum-space electronic structure of graphene-supported ultrathin SnTe.
(a)--(c) Constant-energy maps near the Fermi level for Graphene/SiC, monolayer SnTe, and multilayer SnTe, respectively, measured by momentum microscopy.
In (b) and (c), pseudotetragonal Brillouin zones of SnTe rotated by 0$^\circ$, 30$^\circ$, and 60$^\circ$ are overlaid.
(d) Calculated band structure of 1UC puckered SnTe for the experimentally constrained structure with $a=b=4.49$~\AA\ and a buckling amplitude of 0.13~\AA, with marker size and color representing the Te $p_z$ and $p_x+p_y$ orbital projections.
(e),(h) Monolayer ARPES spectra measured along X--$\Gamma$--X with $p$- and $s$-polarized light, respectively.
(f),(i) Corresponding multilayer spectra.
(g) Experimental geometry for the polarization-dependent ARPES measurements.
}
\end{figure}

Figure~3 shows how this multidomain pseudotetragonal structure appears in momentum space and in the valence-band dispersion. Figures~3(a)--3(c) present constant-energy maps near $E_{\mathrm F}$ for Graphene/SiC, monolayer SnTe, and multilayer SnTe, respectively. In the Graphene/SiC map, multiple Dirac-cone-like contours are observed around $\Gamma$ and are assigned to the well-known replica Dirac cones of graphene on SiC \cite{Polley2019PRB_GrapheneReplicas,Huang2017PRB_MoireGraphene,Nevius2015PRL_SemiconductingGraphene}. Because the same replica features are also visible in the monolayer dataset, we used them as internal references for the momentum calibration of the momentum-microscopy maps.

With this calibration, the monolayer map in Fig.~3(b) shows a set of SnTe-derived hole pockets in addition to the graphene-derived replica features. As expected from the 1UC puckered model in Fig.~1(d), the valence-band maximum lies near the X point, and the superposition of the 0$^\circ$, 30$^\circ$, and 60$^\circ$ domains gives rise to the nearly 12-fold pocket arrangement. By contrast, the multilayer map in Fig.~3(c) retains the same basic momentum-space geometry but becomes much more ringlike, consistent with the broader orientational spread inferred from LEED.

The monolayer spectrum measured with $p$-polarized light [Fig.~3(e)] exhibits a relatively flat band near $E_B \sim 1.0$~eV hybridized with a V-shaped band. Since the valence-band maximum remains near the Fermi level, the experimental valence-band width is about 1.0~eV, smaller than that of the fully relaxed freestanding puckered calculation in Fig.~1(d). We therefore first examined the effect of the experimentally observed lattice compression by fixing the in-plane lattice constants to the LEED value of $a=b=4.49$~\AA\ and performing a structural optimization under this constraint. The resulting compression enhances the buckling amplitude and drives the calculated valence-band width even farther away from the experimental value. We then reduced the buckling amplitude within the same constrained framework and found that a value of 0.13~\AA\ provides the best match to the experimentally relevant valence-band width (see Fig.~S1 in the Supplemental Material \cite{SupplementalMaterial}). 
The recalculated band structure [Fig.~3(d)] agrees well with the ARPES dispersion after aligning the flat-band energy. 

This agreement is not limited to the overall bandwidth. The constrained calculation with a buckling amplitude of 0.13~\AA\ also reproduces the observed polarization dependence: because $p$-polarized light contains an out-of-plane electric-field component, the $p_z$-derived part of the band structure is enhanced in Fig.~3(e), whereas the lower part of the V-shaped band becomes more pronounced with $s$ polarization [Fig.~3(h)], consistent with stronger excitation of in-plane orbitals. Together with the calculations in Fig.~S2 of the Supplemental Material \cite{SupplementalMaterial}, this shows that the flat band is mainly derived from $p_z$ orbitals, the V-shaped band from in-plane orbitals, and that spin--orbit coupling hybridizes them and opens the gap. We therefore conclude that the monolayer film realizes the band structure of a 1UC puckered SnTe layer with a buckling amplitude reduced from the freestanding value, likely due to the substrate effect. 
Because the $\Gamma$--X and $\Gamma$--Y differences are small and ferroelectric domains are mixed within the beam spot, this comparison is used to distinguish between the puckered and nonpuckered models rather than to establish the in-plane anisotropy directly; direct verification of the in-plane anisotropy remains an important subject for future domain-resolved measurements.

For multilayer SnTe, the $p$-polarized ARPES spectrum in Fig.~3(f) exhibits a pronounced ladderlike structure around $\Gamma$, qualitatively consistent with the splitting of relatively flat valence bands by interlayer coupling in the 2UC model of Fig.~1(f). The corresponding $s$-polarized spectrum in Fig.~3(i) emphasizes the more dispersive part of the valence band. In both monolayer and multilayer SnTe, the valence-band maximum lies very close to the Fermi level, indicating strong hole doping.

\begin{figure}[t]
\includegraphics[width=\columnwidth,trim=25 30 530 0,clip]{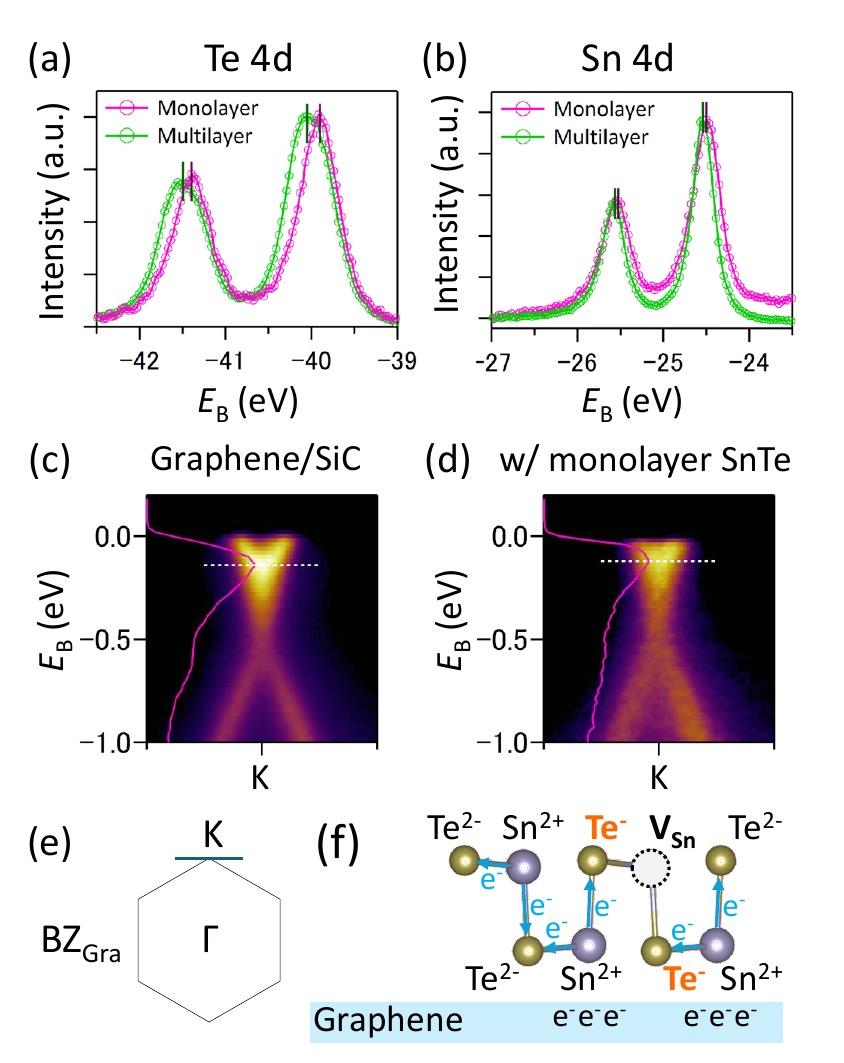}
\caption{
\label{fig:fig4}
Thickness-dependent core-level spectra and graphene-interface signatures in graphene-supported ultrathin SnTe.
(a),(b) Te $4d$ and Sn $4d$ core-level spectra for monolayer and multilayer SnTe.
(c),(d) ARPES spectra near the graphene K point before and after monolayer SnTe growth.
The solid lines indicate EDCs taken at K.
(e) Schematic of the graphene Brillouin zone indicating the K point.
The navy line denotes the momentum cut used for the ARPES measurements in (c) and (d).
(f) Schematic illustration of the proposed interfacial electrostatic picture.
}
\end{figure}

Figure~4 summarizes the thickness-dependent core-level behavior and interface-related signatures in graphene-supported ultrathin SnTe.
In the Te $4d$ spectra [Fig.~4(a)], the multilayer peak shifts to higher binding energy and becomes broader than the monolayer peak.
This trend is consistent with the stronger intrinsic $p$-type character of thicker SnTe, for which Sn vacancies are known to act as acceptor defects \cite{Tan2014JACS_SnTeThermoelectric,Mishra2024JAlloysCompd_SnTeDefects,Hua2021ACSAMI}.
The broader multilayer line shape further suggests a wider distribution of local Te environments, likely associated with defect-affected Te sites and interlayer-related chemical shifts.

By contrast, the Sn $4d$ peak position changes only weakly with thickness, while its linewidth is slightly broader toward the low-binding-energy side in the monolayer [Fig.~4(b)].
This suggests that Sn $4d$ is less sensitive than Te $4d$ to the thickness-dependent Sn-vacancy contribution, while in the monolayer limit it is affected by an additional interface-specific effect, likely enhanced local screening caused by the proximity of graphene.

We also compared the graphene-derived spectra near the K point before and after monolayer SnTe growth [Figs.~4(c)--4(e)].
Within the uncertainty of the present analysis, no clearly resolvable shift of the graphene Dirac cone is established.
This argues against a substantial average charge transfer from graphene into the SnTe film.
Taken together with the Sn $4d$ line-shape broadening, the data instead suggest that the graphene-induced electrostatic effect is more likely a weak local screening at the interface rather than a uniform doping of the SnTe layer as a whole.

The overall picture is summarized in Fig.~4(f): thicker SnTe exhibits a stronger intrinsic $p$-type character, whereas in the monolayer limit the graphene interface mainly affects the local electrostatic environment.
The principal substrate effects supported by the present data are therefore the orientational locking, the renormalization of the buckling amplitude, and a weak interface-specific electrostatic screening in the monolayer limit.

The results of Figs.~2--4 show that graphene is not a passive support for ultrathin SnTe, but actively stabilizes and renormalizes the monolayer limit. Structurally, LEED and momentum microscopy reveal that monolayer SnTe forms discrete domains locked to the graphene lattice, whereas the multilayer already develops a broader, more ringlike orientational distribution. Electronically, comparison between ARPES and DFT shows that the monolayer retains a puckered structure but with a reduced buckling amplitude relative to the freestanding case. In addition, the core-level analysis suggests a weak interface-specific electrostatic screening in the monolayer limit.
Taken together, these results show that graphene controls the orientation and lattice distortion of ultrathin SnTe, while also modifying its local interfacial electrostatic environment.

The comparison between monolayer and multilayer further captures the onset of thickness evolution away from this interface-stabilized puckered limit. The multilayer exhibits a more diffuse momentum-space distribution, interlayer-split valence bands, and a stronger $p$-type character in both valence-band and core-level measurements. These trends indicate that the graphene-induced constraints become weaker with increasing thickness, so that the film begins to recover a more intrinsic SnTe character. In this sense, the monolayer is the interface-stabilized limit, whereas the multilayer marks the beginning of a crossover toward a more bulklike, and possibly more $\beta$-like, state.

In conclusion, we have established the electronic structure of graphene-supported ultrathin SnTe from the monolayer to the multilayer regime. First, we achieved a direct momentum-resolved observation of the electronic structure of monolayer SnTe. Second, the agreement between angle-resolved photoemission spectroscopy, polarization dependence, and first-principles calculations supports the assignment of the experimentally realized monolayer-limit electronic structure to a puckered even-layer biatomic-layer building block rather than a nonpuckered square-flat limit. Third, we showed that graphene actively interacts with ultrathin SnTe by selecting its in-plane orientational texture, renormalizing the buckling amplitude, and modifying the local interfacial electrostatic environment in the monolayer limit. Fourth, by comparing monolayer and multilayer films, we captured the onset of thickness evolution from an interface-stabilized puckered ultrathin limit toward a more bulklike, and possibly more $\beta$-like, state. These results identify graphene-supported ultrathin SnTe as a platform in which puckering, substrate coupling, and thickness evolution can be directly resolved in momentum space.

We thank S. Makita, F. Arai, F. Komori, N. Nagamura, H. Ando, and H. Nomura for helpful discussions. We also thank H. Jin and J. Kim for providing the structural models of SnTe. This work was supported by JSPS KAKENHI Grant No. JP24K01352, JST CREST Grant No. JPMJCR2435, and the Shiraishi Foundation for Science and Development. Part of this work was conducted at Nanofab, Science Tokyo, with support from the Advanced Research Infrastructure for Materials and Nanotechnology in Japan (ARIM), Grant No. JPMXP1225IT0064.

\section*{DATA AVAILABILITY}

The data that support the findings of this article are not publicly available. The data are available from the authors upon reasonable request.


\bibliographystyle{apsrev4-2}
\bibliography{references}

\end{document}


\title{Supplemental Material for\\
``Direct observation of the electronic structure of even-layer puckered SnTe monolayer films on graphene''}

\author{Satoru Ichinokura}
\email{ICHINOKURA.satoru@nims.go.jp}
\affiliation{Center for Basic Research on Materials, National Institute for Materials Science, Tsukuba, 305-0003, Japan}

\author{Shunsuke Tsuda}
\affiliation{Center for Basic Research on Materials, National Institute for Materials Science, Tsukuba, 305-0003, Japan}

\author{Kiyohisa Tanaka}
\affiliation{UVSOR Synchrotron Facility, Institute for Molecular Science, Okazaki, 444-8585, Japan}

\author{Koichiro Yaji}
\affiliation{Center for Basic Research on Materials, National Institute for Materials Science, Tsukuba, 305-0003, Japan}
\affiliation{Unprecedented-scale Data Analytics Center, Tohoku University, Sendai, 980-8578, Japan}
\affiliation{Photon Science Innovation Center (PhoSIC), Sendai, 980-8572, Japan}

\maketitle

\renewcommand{\thefigure}{S\arabic{figure}}

\section{Experimental details}

Ultrathin SnTe films were grown on epitaxial graphene formed on an $n$-type 4H-SiC(0001) substrate. Epitaxial graphene was fabricated by annealing the SiC substrate at 1850$^\circ$C for 15 s in an Ar atmosphere at 800 Torr \cite{Ichinokura2022PRB_LiIntercalatedGraphene, Ichinokura2024ACSNano_CaIntercalatedBilayerGraphene}. After degassing at 560$^\circ$C for 30 min, the graphene substrate was held at 200$^\circ$C, and SnTe was deposited from a Knudsen-cell source charged with SnTe powder \cite{Chang2019APLMat_GrowthPhaseSnTeGraphene}. Because SnTe was evaporated from a compound source rather than from separate elemental sources, Sn and Te were supplied simultaneously from the same Knudsen cell with the nominal stoichiometry of SnTe. The source material was 99.999\% SnTe powder purchased from Kojundo Chemical Laboratory Co., Ltd. All processes from film growth to LEED and photoemission measurements were carried out \textit{in situ} in two separate systems: the LEED/momentum-microscopy apparatus and the UVSOR BL5U ARPES endstation, both equipped with growth chambers.

To prepare well-separated thick-limit and thin-limit SnTe films in the two setups with different source--sample distances, we used the same type of SnTe Knudsen-cell source and 
intentionally adopted sufficiently long and short deposition times in each system.
In the UVSOR BL5U measurements, the thick-limit and thin-limit samples were prepared with deposition times of 120~s and 12~s, respectively, whereas in the LEED/momentum-microscopy measurements the corresponding samples were prepared with deposition times of 3~min and 1~min.

The successful preparation of the thick-limit and thin-limit samples was confirmed independently in the two systems from the mutually consistent LEED patterns and ARPES spectra.
The thick-limit samples exhibited ring-like LEED patterns and multilayer-type ARPES/Fermi-surface features, whereas the thin-limit samples showed the characteristic valence-band dispersion of the monolayer regime.
The thin-limit ARPES spectrum is reproduced well by the 1UC puckered DFT calculation, confirming that the thin-limit sample corresponds to the 1UC SnTe regime.
Accordingly, the monolayer and multilayer regimes in this work refer to these experimentally established thin-limit and thick-limit regimes, respectively.

The in-plane crystallographic orientation, symmetry, and reciprocal-space pattern of the as-grown films were examined by low-energy electron diffraction (LEED) using an electron energy of 74 eV.

Constant-energy maps in momentum space were measured using a Nano-ESCA-based photoemission microscopy apparatus equipped with a 10.9 eV vacuum-ultraviolet laser \cite{Yaji2024eJSSNT_PhotoemissionMicroscopy,Yaji2024STAMMethods_iSPEM,Tsuda2024eJSSNT_ToFPEEM}. The PEEM lens system enabled momentum-space ($k_x$--$k_y$) imaging, and the photoelectrons were energy-filtered with an Imaging Double Energy Analyzer (IDEA). The measurement temperatures were 44 K for the data shown in Fig.~3(a) and Fig.~3(b) in the main text, and 4 K for the data shown in Fig.~3(c).

Valence-band ARPES and core-level photoemission measurements of Te $4d$ and Sn $4d$ were carried out at beamline BL5U of the UVSOR synchrotron facility using an MBS A-1 analyzer \cite{Ichinokura2023APL_hBNAlkaliDoping}. The measurements were performed at 9 K. For valence-band measurements, linearly polarized light with a photon energy of 65 eV was used, and both $p$- and $s$-polarized geometries were employed to distinguish the orbital contributions to the valence bands. For core-level measurements, $p$-polarized light with a photon energy of 175 eV was used.

\section{Computational details}

Electronic-structure calculations were carried out using \textsc{QUANTUM ESPRESSO}~\cite{Giannozzi2009QE,Giannozzi2017QE,Giannozzi2020QE} within the framework of Kohn--Sham density functional theory \cite{Hohenberg1964,KohnSham1965}. Fully relativistic projector-augmented-wave pseudopotentials were used for Sn and Te, and the exchange-correlation functional was treated within the Perdew--Burke--Ernzerhof generalized gradient approximation \cite{PerdewBurkeErnzerhof1996}. The plane-wave cutoff energies for the wave functions and charge density were set to 50 and 250 Ry, respectively. Spin--orbit coupling was included within the noncollinear formalism. Structural relaxation was performed using the Broyden--Fletcher--Goldfarb--Shanno algorithm with a force convergence threshold of $1.0\times10^{-3}$ and an electronic convergence threshold of $1.0\times10^{-6}$, using a $6\times6\times1$ $k$-point mesh. A vacuum spacing corresponding to a $c$-axis lattice parameter of 21.5615~\AA\ was introduced in the slab geometry.

For the spin--orbit-coupled band-structure calculations of the puckered and non-puckered models shown in Fig.~1 of the main text, self-consistent calculations were carried out on a $12\times12\times1$ $k$-point mesh with Methfessel--Paxton smearing of 0.02 Ry, followed by band calculations along the Y--$\Gamma$--X--S high-symmetry line.
For the puckered model [Fig.~1(d),(f)], the lattice parameters were $a=4.560$~\AA, $b=4.530$~\AA, and $c=21.5615$~\AA.
For the non-puckered model [Fig.~1(e),(g)], the in-plane lattice constants were fixed to the average value of the puckered model, namely $a=b=4.545$~\AA.
To retain the inversion-symmetry-broken spin splitting, both \texttt{nosym=.true.} and \texttt{noinv=.true.} were imposed.

For the experimentally constrained calculation shown in Fig.~3(d), the in-plane lattice constants were fixed to the experimental monolayer value, $a=b=4.49$~\AA.

For the orbital-character analysis in Fig.~3(d) of the main text, a separate calculation sequence was employed using a structure constrained to $a=b=4.49$~\AA\ with a fixed buckling amplitude of 0.13~\AA, while only the in-plane atomic positions were relaxed. A subsequent self-consistent calculation was then performed, followed by projection analysis using \texttt{projwfc.x}. Because the spin--orbit-coupled projections are obtained in the $j,m_j$ basis, the resulting complex projection amplitudes were further transformed into out-of-plane $p_z$ and in-plane $(p_x+p_y)$ components using the Clebsch--Gordan decomposition for comparison with polarization-dependent ARPES.

\section{Substrate-Renormalized Buckling of Monolayer SnTe}

\begin{figure}[t]
\centering
\includegraphics[width=\columnwidth]{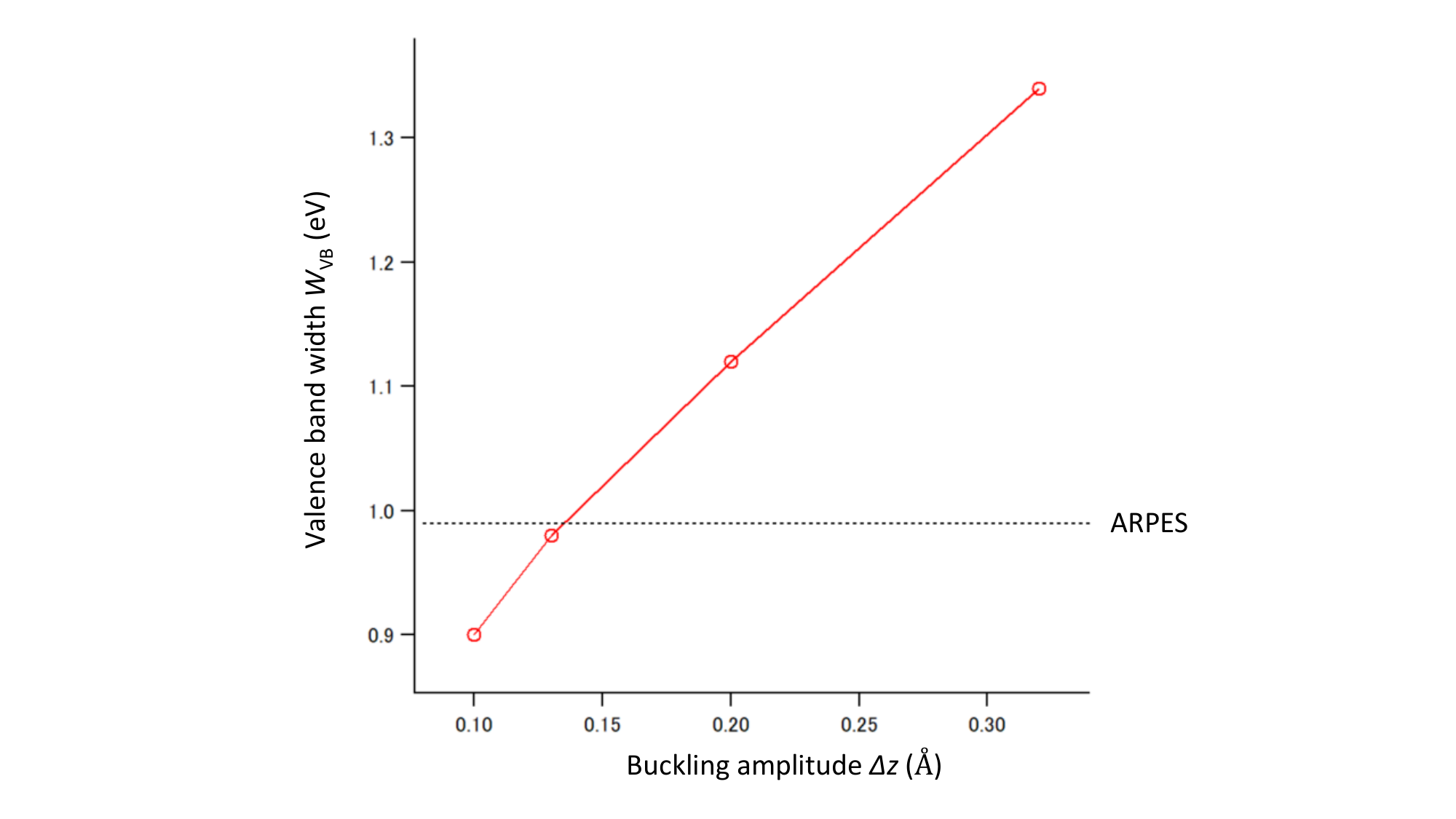}
\caption{
\label{fig:sm_buckling_width}
Calculated valence-band width of monolayer SnTe as a function of the buckling amplitude $\Delta z$, defined consistently with Table~I. The calculations were performed with the in-plane lattice constant fixed to $a=b=4.49$~\AA. The dashed horizontal line indicates the experimental valence-band width of approximately 1.0~eV estimated from ARPES. The closest agreement is obtained for $\Delta z = 0.13$~\AA, which is therefore adopted in the main text as the substrate-renormalized buckling amplitude of monolayer SnTe on graphene.
}
\end{figure}

To further quantify the substrate-limited buckling discussed in the main text, we performed additional DFT calculations while fixing the in-plane lattice constant to the experimental value of $a=b=4.49$~\AA\ and varying the buckling amplitude $\Delta z$ stepwise. Here, $\Delta z$ is defined in the same manner as in Table~I, namely, as the vertical separation within an Sn--Te pair relative to the pair-averaged plane. For each constrained structure, the valence-band width was extracted from the calculated band dispersion in the same manner as in the main text. As shown in Fig.~\ref{fig:sm_buckling_width}, the calculated valence-band width increases monotonically with increasing $\Delta z$. Comparing these values with the experimental bandwidth of approximately 1.0~eV obtained from ARPES, we find that the best agreement is obtained for $\Delta z = 0.13$~\AA. This result supports the interpretation that monolayer SnTe on graphene retains a puckered structure, but with a reduced buckling amplitude compared with the fully relaxed freestanding structure.
The value of $\Delta z = 0.13$~\AA\ should therefore be regarded as a semi-quantitative estimate of the experimentally relevant buckling scale within this constrained Perdew–Burke–Ernzerhof (PBE) framework, rather than as a unique structural determination based solely on the absolute calculated bandwidth.

\section{Orbital Crossing Without Spin--Orbit Coupling}
\begin{figure}[t]
\centering
\includegraphics[width=\columnwidth]{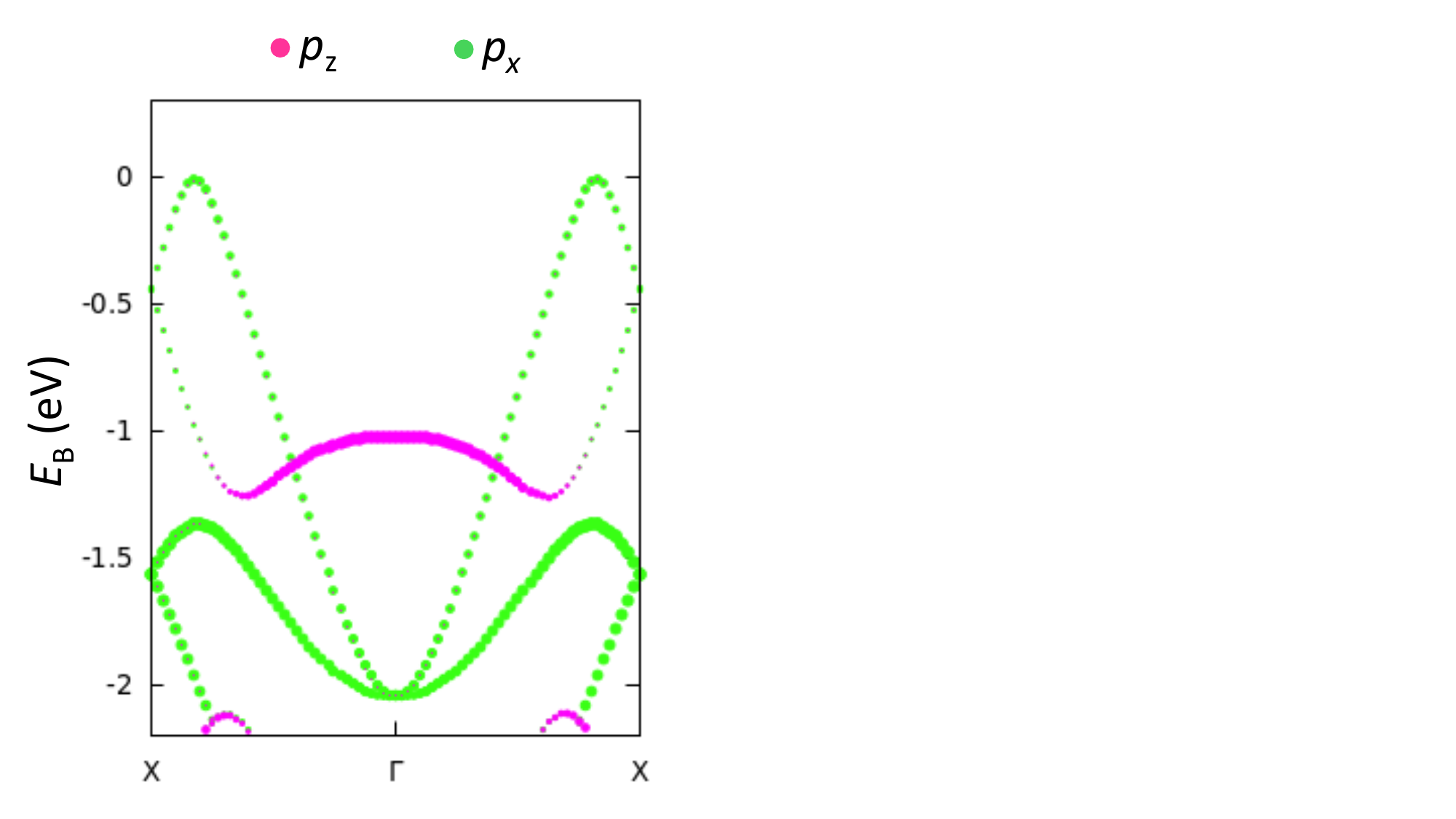}
\caption{
\label{fig:sm_nosoc}
Orbital-projected band structure of monolayer puckered SnTe calculated without spin--orbit coupling for the same constrained structure used in the main text. The marker size indicates the projection onto Te $p_z$ (magenta) and in-plane $p_x$ (green) orbitals. In the absence of SOC, the relatively flat $p_z$-derived band and the dispersive in-plane band cross along X--$\Gamma$--X. In the SOC-included calculation and in the ARPES data shown in the main text, this crossing evolves into an anticrossing, demonstrating SOC-induced hybridization between the two orbital sectors.
}
\end{figure}

To clarify the role of spin--orbit coupling (SOC) in the monolayer band structure, we also calculated the orbital-projected band dispersion without SOC for the same constrained puckered structure with $a=b=4.49$~\AA\ and $\Delta z = 0.13$~\AA\ used in the main text. As shown in Fig.~\ref{fig:sm_nosoc}, the relatively flat $p_z$-derived band and the more dispersive in-plane $p_x$-derived band cross each other along X--$\Gamma$--X when SOC is neglected. In contrast, the SOC-included calculation shown in the main text exhibits a clear avoided crossing between these two bands. The ARPES spectra likewise show the hybridized, anticrossing band structure rather than a simple crossing. This comparison demonstrates that the experimentally observed gap opening is consistently explained by SOC-induced hybridization between the out-of-plane and in-plane orbital components, thereby supporting both the validity of the band-assignment model and the essential role of SOC in graphene-supported monolayer SnTe.

\noindent Reference numbers follow those of the main article.